\documentclass[nofootinbib,aps,notitlepage]{revtex4-2} 

\usepackage{graphicx} 
\usepackage{subcaption}  
\usepackage{amsmath}
\usepackage{amssymb}
\usepackage{latexsym}
\usepackage{xcolor,soul}

\newcommand{\be}{\begin{equation}}
\newcommand{\ee}{\end{equation}}
\newcommand{\bea}{\begin{eqnarray}}
\newcommand{\eea}{\end{eqnarray}}

\begin{document}

\title{Spontaneous Cogensis by QCD axion in Type I Seesaw}

\author{Eung Jin Chun}
\affiliation{Korea Institute for Advanced Study, Seoul 130-722, Korea}

\begin{abstract}
We propose a generic axion-driven cogenesis scenario in which both the baryon asymmetry and dark matter abundance originate from the kinetic misalignment. The framework unifies the Peccei–Quinn (PQ) mechanism with a Type-I seesaw sector, where Hubble-induced masses and higher-dimensional PQ-violating operators drive early-time axion rotation. Working within the DFSZ axion model augmented by heavy neutrinos, we identify the parametric window of right-handed neutrino masses, determined by its decay rate, and the range of Hubble scales compatible with successful cogenesis, while maintaining the axion solution to the strong CP problem and satisfying current limits on axion isocurvature perturbations. Our results establish kinetic axion misalignment as a robust and predictive mechanism for axion cogenesis, independent of the inflationary microphysics.
\end{abstract}

\maketitle

\section{Introduction}
The Peccei–Quinn (PQ) mechanism remains the most compelling resolution of the strong CP problem \cite{pq77, ww78}. It postulates a global symmetry $U(1)_{PQ}$ which, upon spontaneous breaking, gives rise to the axion—a pseudo Nambu–Goldstone boson signaling new physics at high energy scales \cite{ksvz79, dfsz81}.
Beyond addressing the fine-tuning of the QCD vacuum, axions are well-motivated candidates for cold dark matter due to their ultra-light mass and cosmological stability \cite{kc08}. Traditionally, axion dark matter is produced via the standard misalignment mechanism \cite{misal83}, but recent developments have highlighted the kinetic misalignment mechanism \cite{chh19}, where the axion's initial velocity plays a crucial role. 
This kinetic background $\dot{\theta}$ provides a profound link to the baryon asymmetry of the universe \cite{ck87}, opening a pathway to spontaneous baryogenesis \cite{ch19, demy20, axiolepto}.

The high-scale origin of PQ symmetry breaking is plausibly linked to the seesaw mechanism \cite{axion-neutrino83}, providing a unified origin for axions and neutrino masses. In this context, the axion can be identified with a Majoron \cite{cmp80} and mediate spontaneous leptogenesis through both the biasing of chemical potentials during equilibration \cite{cj23} and the out-of-equilibrium decay of right-handed neutrinos \cite{bms17}. The dynamics of such scenarios within the Type-I seesaw framework have been systematically analyzed in Ref.~\cite{cls25a}, with various aspects of spontaneous leptogenesis explored more broadly in Refs.~\cite{ksy14,ik15,berb23,bchp24,wada24,berb25,cddg25,tw26}. A key unresolved issue, however, is the physical origin of the required initial kinetic misalignment. One proposal identifies the PQ field itself as the inflaton, where a pole in its kinetic term triggers axion rotation \cite{hmlee}. While elegant, this construction is highly restrictive: it is essentially limited to KSVZ-type models \cite{ksvz79} and confines the axion decay constant to a narrow range, $f_a < 10^9$ GeV \cite{cls25b}.

In this work, we present a more general and flexible framework in which inflation is decoupled from the PQ sector, yet still governs its early-time dynamics. Focusing on the DFSZ axion model \cite{dfsz81} augmented by right-handed neutrinos, we realize a unified description of PQ symmetry breaking, neutrino mass generation, and axion-driven cogenesis. We first establish the conditions for the successful simultaneous generation of dark matter and baryon asymmetry, identifying the viable range of the right-handed neutrino mass $m_N$ as dictated by its decay rate. We then demonstrate that a Hubble-induced PQ-breaking mass term, together with higher-dimensional PQ-violating operators, naturally generates the required axion kinetic misalignment. Finally, we delineate the allowed parameter space by imposing constraints from the axion quality problem \cite{axionquality} and current limits on axion isocurvature perturbations \cite{planck13}.

\section{DFSZ Axion and Seesaw}
The DFSZ axion model, augmented by the Type I seesaw mechanism, is described by the following Lagrangian:
\begin{eqnarray} \label{Ldfsz}
        -{\cal L}_{\rm DFSZ} &=& y_H \Phi^2 H_u H_d + \frac{1}{2} y_N \Phi N N + h.c. \\
   & +&y_u q u^c H_u +y_d q d^c H_d+y_e l e^c H_d +y_\nu l N H_u + h.c. \nonumber
\end{eqnarray}
where we employ chiral notation for the fermions. Here, $H_u$ and $H_d$ denote the two Higgs doublets that couple to the up-type and down-type fermions, respectively. For simplicity, we suppress flavor indices, assuming flavor independence throughout our analysis.
The PQ charges are assigned as $x_\Phi=1$  and $x_{H_u}=x_{H_c}=-1$. The charges for the remaining fermions are summarized in the following table:
\begin{equation}
    \begin{tabular}{c|c|c|c|c|c|c}
    $\psi$ & ~$q$~ & ~$u^c$~ & ~$d^c$~ &~$l$~& ~$e^c$~ &~$N$~ \\
      \hline
 $x_\psi$  & $+{1\over2}$  & $+{1\over2}$ & $+{1\over2}$ & $+{3\over2}$ &$-{1\over2}$ &  $-{1\over2}$ 
    \end{tabular}
\end{equation}
Upon PQ symmetry breaking, the scalar field $\Phi$ is parameterized as $\Phi = v_a e^{i\theta}/\sqrt{2}$, where $v_a$ is the vacuum expectation value (VEV) and $\theta = a/v_a$ represents the dimensionless axion field. Consequently, the heavy Majorana (right-handed) neutrino $N$ acquires a mass $m_N = y_N v_a/\sqrt{2}$.

Rotating away the axion degree from the above DFSZ Lagrangian, 
it interacts via derivative and anomaly couplings given by
\begin{eqnarray}  \label{Lanomaly}
  - {\cal L}_{\theta} &=& \sum_\psi  x_\psi \partial_\mu \theta \bar{\psi}\bar{\sigma}^\mu \psi +  {c_S \over 32 \pi^2} {\theta} G\tilde{G} + {c_W \over 32 \pi^2} {\theta} W\tilde{W},
\end{eqnarray}
where $\bar{\sigma}_\mu=(1,-\vec{\sigma})$, $c_S$ and $c_W$ are the anomaly coefficients of strong and weak gauge symmetries, respectively.
For $N_f=3$ generations, these coefficients are $(c_S, c_W) = (2N_f, 3N_f)$. Following standard convention, we define the axion decay constant as $f_a = v_a/N_{DW}$, where $N_{DW} = |c_S|$ is the domain wall number, reflecting the degeneracy of the QCD potential, $V(\theta) \sim \Lambda_{QCD}^4 \cos(N_{DW} \theta)$.

\section{Spontaneous Cogenesis}
Assuming that all the interactions in (\ref{Ldfsz}) and (\ref{Lanomaly}) after PQ symmetry breaking reach equilibrium, we obtain the following relations for the chemical potentials, biased by the kinetic background $\dot\theta$ \cite{ch19,demy20}:
\begin{equation}
    \begin{array}{c}
        \mu_q+ \mu_{u^c}+\mu_{{H_u}}=0, \quad
         \mu_l + \mu_{e^c}+\mu_{{H_d}}=0,\\
        \mu_q+ \mu_{d^c}+\mu_{{H_d}}=0,\quad
        \mu_l +\mu_{{H_u}}= x_l \dot\theta,~~~~~\\
        \mu_{H_u}+\mu_{H_d}=(x_{H_u}+x_{H_d}) \dot\theta,\\
         N_f(2\mu_q+  \mu_{u^c} +  \mu_{u^c})=c_s \dot\theta,  \quad 
        N_f(3 \mu_q + \mu_l)=c_w \dot\theta.~~~~~~~~~~
    \end{array}
\end{equation}
The relation in the third line describes the equilibration of the quadratic term $B H_u H_d$ with $B \equiv y_H v_a^2/2$, e.g., via $H_u H_d \to W W$. Given $B \sim (\text{TeV})^2$, this process remains in equilibrium for temperatures $T < 10^7$ GeV. The situation must be contrasted with that of the KSVZ model \cite{cls25b}.  The above equilibrium conditions, combined with the charge neutrality requirement:
\begin{equation}
    N_f(\mu_q - 2 \mu_{u^c} + \mu_{d^c} - \mu_l + \mu_{e^c}) + 2 \mu_{H_u} - 2 \mu_{H_d}=0,
\end{equation}
yield a non-trivial $B-L$ number $n_{B-L}=c_{B-L} \dot\theta T^2/6$ where $c_{B-L} = \left( 8 c_W -16 c_S - 69  x_l \right) \dot\theta/9$. Consequently,  the resulting baryon number is $n_B=c_B \dot\theta T^2/6$ with the coefficient:
\begin{equation}
    c_B = {28 \over 79} {1\over 9}(8 c_W-16 c_S-69 x_l).
\end{equation}
Using the values $c_W=3N_f$, $c_S=2N_f$ and $x_l=3/2$, we find $c_B=-2.3 $.
This solution is derived under the assumption that the decay and inverse decay processes of right-handed neutrinos are in equilibrium, which holds when the neutrino Yukawa coupling $y_\nu$ is sufficiently large \cite{cj23}. In the opposite regime, one must incorporate the full Boltzmann evolution to properly account for the $\dot\theta$-driven asymmetric decay of $N$.

Following the analysis in \cite{cls25a}, 
the final baryon asymmetry $Y_B=n_B/s$ can be parameterized as
\begin{equation} \label{eqYB}
    Y_B= \kappa(K) |c_B| \left({{1\over6} \dot\theta T^2 \over s }\right)_{T=m_N}
\end{equation}
where the observed value is $Y_B=0.87 \times 10^{-10}$.
In the above equation, $\kappa(K)$ denotes the efficiency factor as a function of $K\equiv \Gamma_N/H_1$ where $\Gamma_N= y_\nu^2 m_N/8\pi$ is the decay rate of $N$ and $H_1$ is the Hubble parameter calculated at $T=m_N$.  The values for $\kappa(K)$ are adopted from Fig. 3 of \cite{cls25a}.

\medskip

Given the kinetic misalignment $\dot\theta$, the axion abundance $n_a/s$, where $n_a=\dot\theta v_a^2$ is the axion number density and $s=2\pi^2 g_* T^3/45$ is the entropy density, can account for the required dark matter abundance, provided $m_a n_a/s=0.22$ eV \cite{ch19}.  This requires the conserved quantity $n_a/s$  to be
\begin{equation}
    {n_a \over s} = 38.6 \left( {f_a \over 10^9 \mbox{GeV} } \right),
\end{equation}
where we have used the relation $m_a=5.7 \mbox{meV} (10^9\mbox{GeV}/f_a)$ \cite{axionmass16}. 
By combining this with Eq. (\ref{eqYB}), we find that successful cogenesis is achieved for a right-handed neutrino mass of
\begin{equation}
    m_N=3.7  {N_{DW} \over \sqrt{|c_B| \cdot \kappa(K)} }  \left( f_a \over 10^9 \mbox{GeV} \right)^{1/2} \mbox{TeV}.
\end{equation}
In Fig.~1, we illustrate the allowed values of $m_N$ as a function of $K$ for two distinct initial abundance scenarios for $N$. The thermal initial condition $Y_N(0) = Y_N^{\text{eq}}$ may be realized if additional interactions for $N$ are present in the early universe. Let us remark that our analysis for the DFSZ axion model is applied to the allowed range of the axion decay constant $f_a\approx [2.5\times 10^8, 10^{12}]$ GeV where the lower bound comes from SN1987A data \cite{cr24}.

\begin{figure}[!ht]
    \centering
\includegraphics[width=0.45\linewidth]{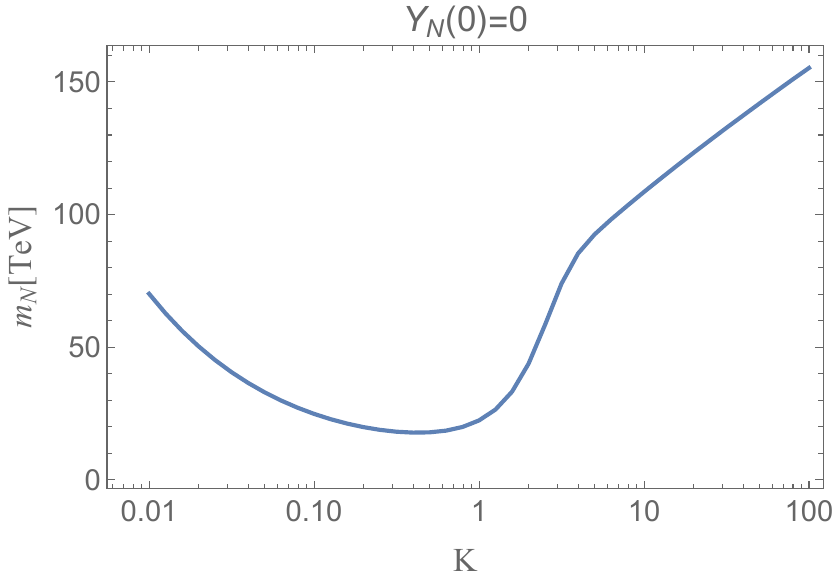} 
\hspace{1.5em}
\includegraphics[width=0.45\textwidth]{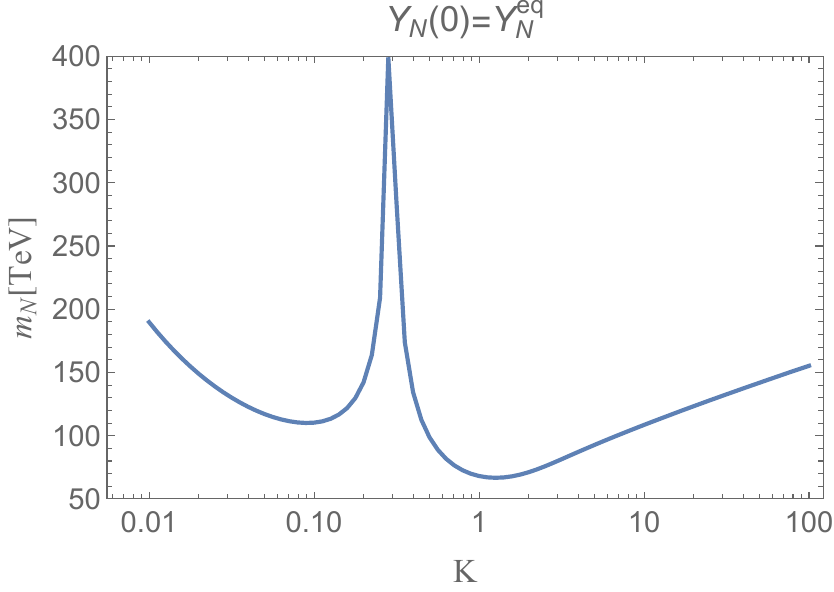}
    \caption[short]{ 
   The required right-handed neutrino  mass $m_N$ as a function of the decay parameter $K$ for two distinct initial conditions: $Y_N(0)=0$ (vanishing initial abundance) and $Y_N(0)=Y_N^{\rm eq}$ (thermal initial abundance). For this analysis, we have adopted the benchmark values $N_{DW}=6$, $|c_B|=2.3$, and $f_a=10^9$ GeV.}
    \label{fig:mN-K}
\end{figure}

\section{Hubble-induced kinetic misalignment}

To model PQ symmetry breaking, we consider a typical scalar potential 
\bea \label{VPQC}
V_{\rm PQC}=-(\mu_\Phi^2+a_H H^2) |\Phi|^2 + \lambda_\Phi |\Phi|^4 .
\eea
The inclusion of the Hubble-induced mass term ($a_H H^2$) \cite{drt95,ejc14} ensures that the PQ symmetry remains broken throughout the universe's expansion and thus there is no domain wall problem.  The field $\Phi$ is parameterized in terms of its radial and axial modes as follows:
\begin{equation} \label{phivev}
    \Phi= {\phi \over \sqrt{2}} e^{i\theta} 
    ~~\mbox{with}~~ \theta={a \over \phi}
    ~~\mbox{and}~~
    \phi=v_a \sqrt{1+ {H^2\over H_c^2}}.
\end{equation}
Here, $ v_a= \mu_\Phi/\sqrt{\lambda_\Phi}$ and $H_c=v_a \sqrt{\lambda_\Phi/a_H}$.
The kinetic background of the axial mode $\theta$ is supposed to generate the observed baryon asymmetry before eventually trasitioning into cold dark matter as described in the previous section.  
To trigger kinetic misalignment, we introduce a PQ symmetry violating (PQV) term  \cite{ad85} that is provided by a higher-dimensional operator:
\begin{equation} \label{VPQV}
    -V_{\rm PQV}= \lambda_n\frac{\Phi^n}{M_P^{n-4}}+\text{h.c.} ={|\lambda_n| \over \sqrt{2^n}}
    {\phi^{n}\over M_P^{n-4}} 2\cos(n\theta+\delta_n) .
\end{equation}
The presence of such a term may be guaranteed by a discrete gauge symmeetry \cite{cl92}. 
We require that  the operator is sufficiently suppressed so as not to disturb the radial dynamics of $\phi$, and  compromise the axion’s resolution of the strong CP problem~\cite{axionquality}.

While the radial mode follows the evolution in Eq. (\ref{phivev}), the axial mode $\theta$ evolves according to the equation of motion:
\begin{equation}
    \ddot{\theta}+ (3H + 2 {\dot\phi \over \phi}) \dot\theta + {1\over \phi^2} {\partial V_{\rm PQV} \over \partial \theta}=0.
\end{equation}
We assume an initial misalignment $\theta_I \neq 0$ is established during inflation. Following inflation, the inflaton field  is supposed to oscillate coherently until the universe reheats.
\begin{figure}[!ht]
    \centering
    \includegraphics[width=0.55\linewidth]{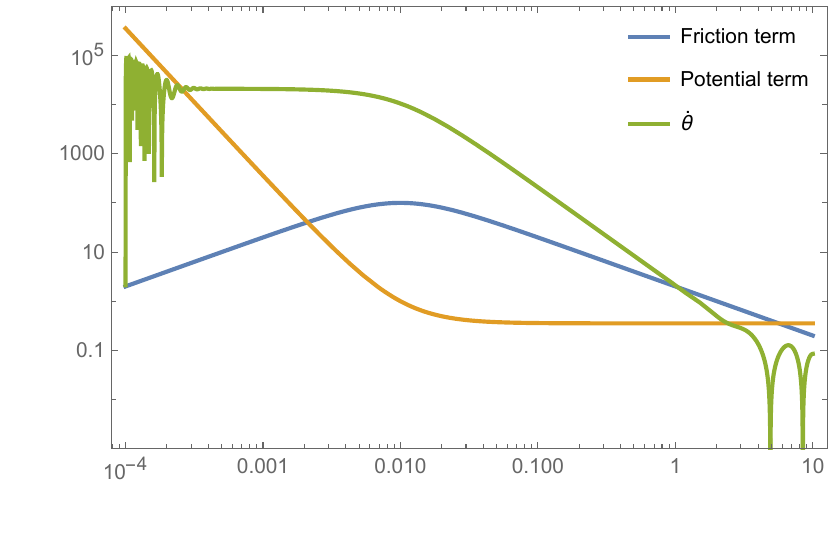}
    \caption[short]{ 
    Evolution of the friction, potential term and $\dot\theta$ in terms of the time $t$ with arbitrary initial conditions and normalization setting $t_c=0.01$.}
    \label{fig:gamma}
\end{figure}
Our analysis focuses on the matter-dominated era ($H=2/3t$), extending from the end of inflation to the onset of radiation domination ($H=H_R$). The axion dynamics are explicitly governed by
\begin{equation}
\begin{split}
    \ddot{\theta}+{2\over t} \left(1- {t_c^2/t^2 \over 1+t_c^2/t^2 }\right) \dot{\theta} +{1\over n} \delta m_a^2(t) \sin(n\theta+\delta_n)=0 ~~~~~~~~ \\
    \delta m_a^2(t)\equiv\delta m_{a0}^2 \left(1+{t_c^2\over t^2}\right)^{n-2},\quad
    \delta m_{a0}^2\equiv2n^2 {|\lambda_n| \over \sqrt{2^n}} {v_a^{n-2} \over M_P^{n-4}}
\end{split}
\end{equation}
where $t_c$ is the time-scale corresponding to $H_c$.
Figure 2 illustrates the interplay between the friction term, the potential term, and $\dot{\theta}$. Initially, the kinetic mode $\dot{\theta}$ grows and oscillates until it becomes comparable to the potential term: $\dot{\theta}(t) \approx \delta m_a(t)/\sqrt{n}$ at time $t_0$ (or $H_0$). We utilize these values at $t_0$ as our effective initial conditions, which are set by specific inflationary model parameters.
Notably, as $\delta m_a(t) \propto 1/t^{n-2}$ decays rapidly, $\dot{\theta}$ stabilizes while $\phi^2$ evolves as $1/t^2$. This results in the conservation of the axion number density $n_a(t) = \dot{\theta}(t) \phi^2(t)$ in a volume:
\begin{equation}
   n_a(t)= \dot\theta_0 \phi^2_0 (t_0/t)^2~~ \mbox{with} ~~
\dot\theta_0=\delta m_a(t_0)/\sqrt{n}~~ \mbox{and} ~~\phi_0^2=v_a^2 (1+H_0^2/H_c^2).
\end{equation}
This behavior contrasts with thermal mass correction models where $\phi^2 \propto 1/t$ and $\dot{\theta} \propto t^{1/2}$ during the radiation era \cite{chun24}.

Upon entering the radiation-dominated era, the axion-to-entropy ratio $Y_a = n_a/s$ becomes a conserved quantity. At the reheating temperature $T_R$, this is expressed as:
\begin{equation}
    Y_a = {\dot\theta_0 \phi_0^2 \over s(T_R) }{H_R^2 \over H_0^2} 
    ~~\mbox{where}~~ 3  H_R^2 M_P^2= {\pi^2\over30} g_* T_R^4. 
\end{equation}
\begin{figure}[!ht]
    \centering
\includegraphics[width=0.7\linewidth]{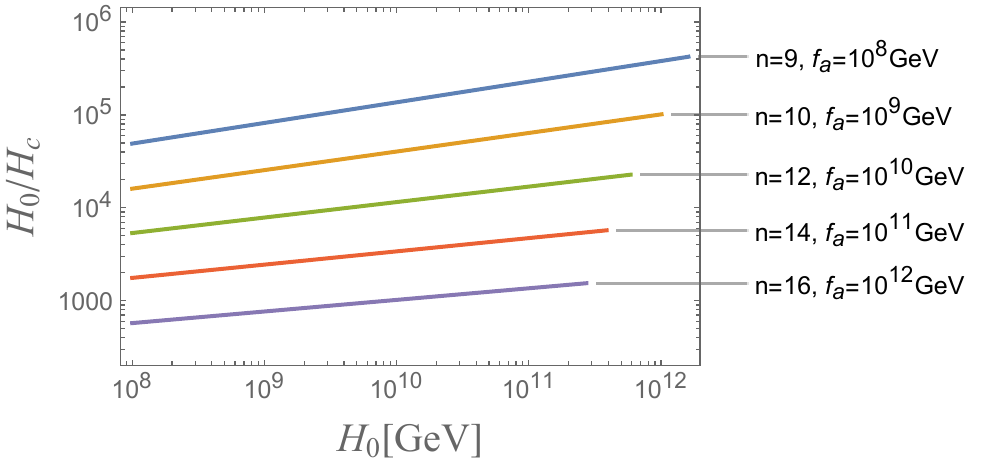}
    \caption[short]{ 
    Cogenesis contours are shown for various choices of $(n,f_a)$.  For each value of $f_a$, the dimensionality $n$ of the PQV operator is selected to preserve axion quality, while the upper bound on $H_c$ is determined by constraints from axion isocurvature perturbations. 
    The figure corresponds to $T_R=10^9$ GeV, although the results are not significantly sensitive to this choice. 
    }
    \label{H0Hc}
\end{figure}
Figure 3 shows the viable parameter space for $(H_0, H_c)$ across various $(n, f_a)$ combinations, taking $N_{DW}=6$, $|\lambda_n|=1$, and $T_R=10^9$ GeV. The results are robust against variations in $|\lambda_n|$ and $T_R$ due to their weak power-law dependence: $H_c \propto |\lambda_n|^{1/2n} T_R^{1/n}$.  To ensure the axion effectively solves the strong CP problem, we required $\delta m_{a0}^2 < 10^{-10} m_a^2$ to obtain $n$ corresponding to $f_a$. 
In our framework, the PQ symmetry is already broken during inflation, which inevitably leads to the generation of axion isocurvature perturbations given by $\delta \theta = H_I / (2\pi \phi_I)$. Here, $H_I$ denotes the Hubble parameter during inflation, and $\phi_I \approx v_a H_I / H_c$ (with $v_a = N_{DW} f_a$) represents the corresponding field value of the PQ scalar. Observations of the cosmic microwave background require the isocurvature power spectrum to be suppressed to within a few percent of the adiabatic scalar power spectrum \cite{planck13}, imposing the following constraint  \cite{ejc14}:
\begin{equation}\left( \frac{H_I}{2\pi \phi_I} \right)^2 \left( \frac{f_a}{10^{12} , \text{GeV}} \right)^{7/6} \lesssim 2.2 \times 10^{-11}.
\end{equation}
This condition translates into the upper bounds on $H_c$ for given $f_a$ indicated by the endpoints in Fig. 3. Our findings demonstrate that cogenesis remains achievable across a broad range of $H_0<H_I$ and $H_c < f_a$ values. This provides a robust and viable pathway for the simultaneous generation of baryon asymmetry and dark matter within a  PQ framework unified with a seesaw mechanism. 

\section{Conclusion}

In this work, we have presented a comprehensive unified model for dark matter and baryogenesis based on DFSZ axion dynamics and the Type-I seesaw mechanism. By employing a Hubble-induced mass term, we ensure that the PQ symmetry is broken during inflation and remains so throughout the cosmic evolution, naturally circumventing the domain wall problem. The kinetic misalignment of the axion, initiated by higher-dimensional PQ-violating operators, provides the necessary $\dot{\theta}$ background to drive spontaneous leptogenesis. Our analysis identifies a robust parameter space where the observed baryon asymmetry and dark matter abundance are concurrently satisfied. 
Specifically, for an axion decay constant $f_a \in [2.5 \times 10^8, 10^{12}]$ GeV, successful cogenesis predicts a right-handed neutrino mass $m_N=[{\cal O}(10), {\cal O}(100)](f_a/10^9 \mbox{GeV})^{1/2}$ TeV within a reasonable range of the decay rate.
We have also accounted for the stringent constraints imposed by axion isocurvature perturbations and the axion quality problem, showing that the model remains viable in a wide range of parameter space.  The interplay between the PQ field mass expressed in terms of the Hubble parameter ($H_c$) and the initial misalignment at $H_0$ provides a flexible yet predictive framework allowed for a broad range of $f_a$ values, establishing a consistent link between the strong CP problem, neutrino physics, and the dark sector of the universe.  Although our primary focus is on the DFSZ framework, the analysis can be readily extended to the KSVZ model. Furthermore, exploring its application to alternative seesaw mechanisms presents an intriguing direction for future research.

\section*{Acknowledgments}
We acknowledge support from APCTP under the program “APCTP-2026-F02.”

\end{document}